\documentclass{emulateapj}
\usepackage{graphicx}
\usepackage{natbib}
\newcounter{Rco}
\newcommand{\Ionst}[1]{\setcounter{Rco}{#1}\Roman{Rco}}
\newcommand{\Ion}[2]{\mbox{#1\ {\scriptsize\Ionst{#2}}}}
\newcommand{\Ionw}[3]{\mbox{#1\ {\scriptsize\Ionst{#2}}~$\lambda\,#3$\,\AA}}

\newcommand{\Ionww}[3]{\mbox{#1\ {\scriptsize\Ionst{#2}}~$\lambda\lambda\,#3$\,\AA}}

\newcommand{\loggw}[1]{\mbox{$\log g\hspace{-0.5mm} =\hspace{-0.5mm}  #1$}}

\newcommand{\ab}[1]{\mbox{Figure\,\ref{#1}}}
\newcommand{\aba}[1]{\mbox{Figure\,\ref{#1}}}
\newcommand{\sA}[1]{\mbox{(Figure\,\ref{#1})}}

\newcommand{\se}[1]{\mbox{Sect.\,\ref{#1}}}
\newcommand{\sK}[1]{\mbox{(Sect.\,\ref{#1})}}
\newcommand{\sla}{\raisebox{-0.10em}{$\stackrel{<}{{\mbox{\tiny $\sim$}}}$}}

\newcommand{\ta}[1]{\mbox{Table\,\ref{#1}}}
\newcommand{\sT}[1]{\mbox{(Table\,\ref{#1})}}
\newcommand{\Teff}{\mbox{$T_\mathrm{eff}$}}
\newcommand{\Teffw}[1]{\mbox{$\Teff\hspace{-0.5mm}=\hspace{-0.5mm}#1\,\mathrm{kK}$}}
\newcommand{\TeffwM}[1]{\mbox{$\Teff\hspace{-0.5mm}=\hspace{-0.5mm}#1\,\mathrm{MK}$}}
\newcommand{\vsgr}{\object{V4743\,Sgr}}
\def\tmap{\emph{TMAP}}

\begin{document}

\title[NLTE model atmospheres for the hottest white dwarfs]
      {NLTE model atmospheres for the hottest white dwarfs:
       Spectral analysis of the compact component in nova \vsgr\thanks{Based on observations collected with \emph{XMM-Newton}, 
       an ESA Science Mission with instruments and contributions directly funded by ESA Member States and the USA (NASA).}
      }

\author{T\@. Rauch$^1$,
        M\@. Orio$^{2, 3}$,
        R\@. Gonzales-Riestra$^4$,
        T\@. Nelson$^{5, 6}$,\\
        M\@. Still$^{7, 8}$,
        K\@. Werner$^1$,
        J\@. Wilms$^9$              
       }

\address{$^1$ Institute for Astronomy and Astrophysics,
              Kepler Center for Astro and Particle Physics,
              Eberhard Karls University, 
              Sand 1,
              D-72076 T\"ubingen, 
              Germany
        }
\address{$^2$ Istituto Nazionale di Astrofisica,
              Osservatorio Astronomico di Padova, 
              vicolo Osservatorio 5, 
              I-35122 Padova, Italy
        }
\address{$^3$ Department of Astronomy, 
              University of Wisconsin-Madison, 
              475 N\@. Charter Street, 
              Madison, 
              WI 53706, 
              USA
        }
\address{$^4$ \emph{XMM-Newton} Science Operations Centre, 
              European Space Astronomy Centre (ESAC), 
              P\@. O\@. Box 78, 
              Villanueva de la Ca\~nada,
              28691  Madrid,
              Spain
        }

\address{$^5$ NASA Goddard Space Flight Center, 
              Greenbelt, 
              MD 20771, 
              USA
        }
\address{$^6$ University of Maryland, 
              Baltimore County, 
              1000 Hilltop Circle, 
              Baltimore, 
              MD 21250, 
              USA
        }
\address{$^7$ NASA Ames Research Center,
              M/S 244-30,
              Moffett Field, 
              CA 94035,
              USA 
        }
\address{$^8$ University College London,
              Mullard Space Science Laboratory,
              Holmbury St Mary,
              Dorking,
              Surrey RH5 6NT,
              United Kingdom 
        }
\address{$^9$ Dr\@. Remeis-Sternwarte, 
              Astronomical Institute of the University Erlangen-Nuremberg,  
              Sternwartstr\@. 7, 
              D-96049 Bamberg, 
              Germany
        }
\email{rauch@astro.uni-tuebingen.de, orio@astro.wisc.edu}

\begin{abstract}
          Half a year after its outburst in September 2002, nova \vsgr\ evolved into the brightest 
          supersoft X-ray source in the sky with a flux maximum around 30\,\AA.
          We calculated grids of synthetic energy
          distributions (SEDs) based on NLTE model atmospheres
          for the analysis of the hottest white dwarfs
          and present the result of fits to 
          \emph{Chandra} and \emph{XMM-Newton} grating X-ray spectra
          of \vsgr\ of outstanding quality, exhibiting
          prominent resonance lines of
          \Ion{C}{5}, \Ion{C}{6}, \Ion{N}{6}, \Ion{N}{7}, and \Ion{O}{7} in absorption.
          The nova reached its highest effective temperature (\Teffw{740\pm 70}) around April 2003 and 
          remained at that temperature at least until September 2003. 
          We conclude that the white dwarf is massive, $\approx 1.1 - 1.2\,\mathrm{M_\odot}$. 
          The nuclear-burning phase lasted for 2 to 2.5 years after
          the outburst, probably the average duration for a classical nova.
          The photosphere of \vsgr\ was strongly carbon deficient ($\approx 0.01$ times solar) and 
          enriched in nitrogen and oxygen ($>5$ times solar). Especially
          the very low C/N ratio indicates that the material at the white dwarf's surface
          underwent thermonuclear burning. 
          Thus, this nova retained some of the accreted material and did not eject all of 
          it in outburst. From March to September 2003, the nitrogen abundance is strongly
          decreasing, probably new material is already been accreted at this stage.\\

\noindent{\it Key words\/}: stars: abundances --
                            stars: AGB and post-AGB --
                            stars: atmospheres --
                            stars: individual: \vsgr\ --
                            novae, cataclysmic variables --
                            white dwarfs

\end{abstract}


\maketitle

\section{Introduction}
\label{sect:intro}

Nova \vsgr\ (\object{Nova Sgr 2002 c}) was discovered in outburst in September
2002 and reached $V=5$\,mag on 2002 September 20 \citep{HEA02}.
It was a very fast nova, with a steep decline in the optical light curve and large ejection
velocities. The time to decay by 3\,mag in the visual (t$_3$), was 15 days and the
FWHM of the H$\alpha$ line reached 2\,400\,km\,sec$^{-1}$
\citep{KEA02}. Estimates of the distance from infrared observations vary from
$1\,200 \pm 300$\,pc \citep{NS03} to $\approx 6\,300$\,pc (Starrfield \& Lyke priv\@. comm.).

In December 2002, the nova was observed for the
first time with \emph{Chandra} ACIS-S. 
At that time it was a very soft and moderately luminous X-ray
source, with a count rate of about 0.3 cts sec$^{-1}$. There were indications that
it was not at the peak of X-ray luminosity yet,
so a  \emph{Chandra} HRC+LETG observation was done only later, on
2003 March 20. 
The count rate was astonishingly high, 40 cts sec$^{-1}$  during 3.6 hours,
then a slow decay, lasting for an hour and a half, was followed by
another 1.5 hours of very low luminosity with  a measurement of only
$\approx 0.02$ cts sec$^{-1}$ \citep{NEA03, SEA03}.
This decline was not due to an eclipse,
since the orbital period of the system is $24\,278 \pm 259$\,sec 
\citep[or 6.74 hours,][]{W8176}
and no eclipse was observed in a following 10\,h \emph{XMM-Newton} observation
 two weeks later.  
An obscuration of a supersoft X-ray source in a nova was
observed once before with BeppoSAX, in \object{V382\,Vel} \citep{OEA02}. The
only reason for  this sudden nearly shut down of the source
may have been the ejection of a new shell of material that was
optically thick to supersoft X-rays from the surface \citep{O08}.
During the \emph{Chandra} observation in March of 2003 an oscillation with a period of
1\,324\,sec was detected, with fluctuations of 20\% from the mean
count rate \citep[see][]{SEA03, NEA03}.
The most likely cause of this oscillation is a non-radial pulsation
of the WD \citep{SEA03}.

A second observation was proposed by us to the \emph{XMM-Newton}
Project Scientist as a target during the Discretionary Time. The nova was
observed for 10\,h with this satellite on 2003 April 4 
\citep{OEA03}.
Three X-ray telescopes with five X-ray detectors were all used:
the European Photon Imaging Camera pn \citep{SEA01}, two  EPIC MOS cameras 
\citep{TEA01}, and two Reflection
Grating Spectrometers \citep[RGS-1 and  RGS-2,][]{HEA01}. The observation lasted a
little over 36\,000\,sec.
The data obtained by the two  EPIC MOS cameras suffered from very severe pile up,
but the operation of the pn camera was switched to timing mode 
and yielded a measured
average, background corrected EPIC-pn count rate 
of $1\,348.0 \pm 0.3$ cts sec$^{-1}$ \citep{OEA03}, with variations 
by 40\% \citep{LEA06}. The RGS count rates were about 57 cts sec$^{-1}$, and 
the unabsorbed flux was $1.5 \times 10^{-9}$ erg cm$^{-2}$ sec$^{-1}$,
consistent with the flux measured 16 days earlier with \emph{Chandra}
\citep{LEA06}.
Most of the variability is well represented as a combination of oscillations at a set of
discrete frequencies lower than 1.7\,mHz \citep{LEA06}. At least five frequencies
preserve their coherence over the 16 days time interval
between the two observations. The 1\,324\,sec period detected with
\emph{Chandra} could be split into two periods, of 1\,310.1\,sec and 1\,371.6\,sec,
both of which are consistent in the \emph{Chandra} data but it may not have been
resolved \citep{LEA06}.  
In that article we suggested that a period in the power spectrum of
both light curves at the frequency of $\simeq$0.75\,mHz
(corresponding to 1\,371.6\,sec) is the spin period of the white
dwarf in the system, and that other observed frequencies
are signatures of non-radial white dwarf pulsations.

The X-ray evolution was followed with 10\,000\,sec long exposures with
the \emph{Chandra} LETG in July and September 2003 and in February 2004,
and with a 30\,000\,sec observation with \emph{XMM-Newton} in September 2004. 
By this time, the nova X-ray luminosity had decayed by 5 orders of magnitude, but
the period of $\approx 1\,370$\,sec was still detectable.
 However, the S/N of the RGS spectra at this epoch was very poor.

We have performed a NLTE spectral analysis of the grating spectra obtained between
 2003 March and 2004 February, with NLTE model-atmosphere techniques. The \emph{TMAP} model-atmosphere
code and the construction of the model atoms that are used are described in \se{sect:models}. An attempt to
determine photospheric parameters with a \emph{XSPEC} fit procedure is presented in \se{sect:preliminary}.
This is followed by a detailed investigation of the observed, flux-calibrated spectrum of \vsgr\ \sK{sect:fine}.
In \se{sect:spevolution}  we describe how our models fit the \emph{Chandra} and
\emph{XMM-Newton} observations to interpret the post-nova evolution in the 18 months after the
 outburst of \vsgr.

\section{The use of model atmospheres}
\label{sect:useofmodels}

 A previous analysis of \emph{Chandra} LETG-S observations of \vsgr\ \citep{PEA05}, 
 done with the \emph{PHOENIX} NLTE models with solar abundances, reached \Teffw{580}. 
 The fit of the model to the data was still poor, and this is not surprising because 
 in the phase directly after the outburst, i.e\@. when the H burning is still on-going 
 (for years), the surface composition of the WD is poorly known, but it is very unlikely 
 to be solar. With the \emph{PHOENIX} code, the nova atmosphere is approximated as an
 expanding, but stationary in time structure. 
 Recently, \citet{VRN10} presented a new version of \emph{PHOENIX} that considers mass-loss 
 as well as velocity fields. 
 In the future, such codes will become a powerful tool for the analysis and understanding 
 of novae and other supersoft sources. 

 However, in order to make progress, we decided to neglect the velocity field, and to 
 use our static NLTE models to investigate basic parameters like
 \Teff\  and surface composition. 
 This is not fully justified because
 especially in the first spectra the lines were significantly blue-shifted,
 but  our aim is at
 least a qualitative modeling of the \emph{XMM-RGS} and \emph{Chandra-LETG} observations
 done between 2003 March and 2004 February. We started with 
 the highest S/N spectrum, the one obtained with the \emph{XMM-Newton}
 RGS gratings on April 4 2003, that we used as a template to adjust
 the abundances. We then proceeded to fit also the \emph{Chandra} spectra
 with the same model, 
 checking whether the abundances we obtained were suitable, and
 adjusting \Teff\ for each epoch. In this
 way, we obtained the evolution of \Teff,
 and the duration of the constant bolometric luminosity phase.

\begin{figure}[ht]
\begin{center}
  \resizebox{\hsize}{!}{\includegraphics[]{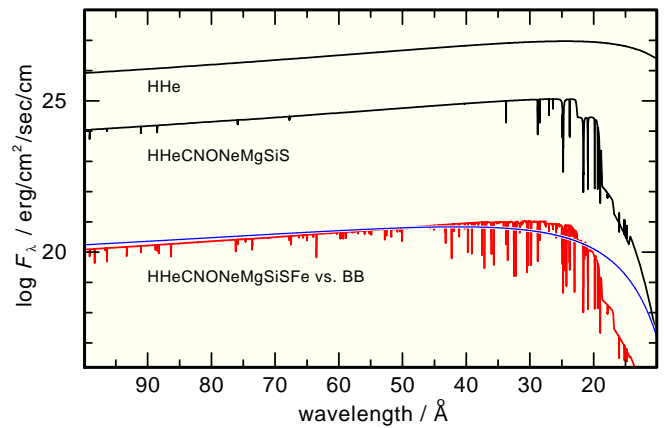}}
  \caption[]{Comparison of synthetic spectra (\Teffw{700}, \loggw{9}),
             from models
             with different elemental composition (see labels). 
             The abundances are given in \ta{tab:abundances} (model A). 
             For clarity, the HHe and the H-Fe spectra are shifted in
             $\log F_\mathrm{\lambda}$ by $+2$ and $-4$, respectively.
            }
  \label{fig:ionselect}
\end{center}
\end{figure}

\subsection{Model atmospheres and atomic data}
\label{sect:models}

For a reliable analysis of X-ray observations of the hottest white dwarfs, detailed NLTE model-atmospheres
that consider opacities of all elements from hydrogen up to the iron-group elements \citep[cf\@.][]{R03}
are required. Thus, for our analysis, we employed the plane-parallel, static models
calculated with the T\"ubingen NLTE Model-Atmosphere Package 
\citep[\tmap\footnote{http://astro.uni-tuebingen.de/$^\sim$rauch/TMAP/TMAP.html},][]{WEA03}.
The construction of model atoms which are used within \tmap\ follows \citet{RD03}. 
Some details for these extremely hot model atmospheres in the MK-\Teff\ range are summarized briefly in the
following.

Since the surface composition is unknown, we started with the calculation of exploratory H+He+C+N+O models and
later included Ne, Mg, Si, S, as well as the iron-group elements. The iron-group elements (Sc -- Ni)
and Ca are treated with a statistical method \citep{RD03} and are represented by one generic model atom. 

Beside an element selection, also the construction of the model atoms has to be performed with care.
An unrealistic upper cut-off in the number of considered iron-group ionization stages causes an 
artificial over-population of the highest ionization stage,
and thus, affects its lines and the flux level. 
In \ab{fig:ionselect}, the impact of iron-group opacities on the astrophysical flux is demonstrated.

\begin{figure}[ht]
\begin{center}
  \resizebox{\hsize}{!}{\includegraphics[]{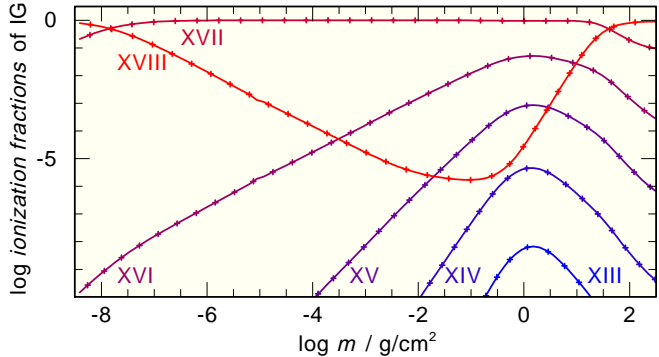}}
  \caption[]{Ionization fractions of the generic iron-group (IG) model atom in a \Teffw{700}, \loggw{9} model. 
            \Ion{IG}{17} is dominant in the line-forming region ($-5.6 < \log m < 0.5$). 
            }
  \label{fig:ionfrac}
\end{center}
\end{figure}

For all values of \Teff, the necessary ionization stages of all atoms are determined in advance by test calculations. 
The selection criterion is,
that at least the ion (e.g\@. in the case of species X), X$^{\mathrm{n}+}$ that is dominant in the line-forming region 
has to be included together with the neighboring two, 
i.e\@. X$^{\mathrm{(n-1)}+}$ and X$^{\mathrm{(n+1)}+}$. Thus, the model atoms 
contain in general three to five ionization stages. E.g\@. in the case of our generic iron-group model atom,
\Ion{IG}{17} is dominant at \Teffw{700} and \loggw{9} [cm sec$^{-2}$] 
\sA{fig:ionfrac}. We therefore selected the ionization stages {\sc xiv} -- {\sc xviii}.
Statistics of the model atoms are shown in \ta{tab:stat}. In total, 228 atomic levels are treated in NLTE, 360 additional
levels in LTE, and 349 individual line transitions are considered.

\begin{table}[ht]
\caption{Statistics of the model atoms used in our calculations of the NLTE model atmospheres with \Teff\ = 0.7\,MK. 
  The notation is: N = levels
  treated in NLTE, L = LTE levels, R = radiative bound-bound transitions. For the generic iron-group (IG) model atom
  (which represents Ca, Sc, Ti, V,  Cr, Mn, Fe, Co, and Ni) the number of individual lines which are combined to 
  so-called superlines is given in brackets.
  }
\label{tab:stat}
\begin{tabular}{lrrr@{\hbox{}\hspace{20mm}\hbox{}}lrrrr}
\hline
\hline
\noalign{\smallskip}
ion & N & L & R & ion & N & L & R \\
\hline
\noalign{\smallskip}
\Ion{H}{1}  &   5  &  11  &  10 & \Ion{Ne}{7}  &  10  &  50  &  12 &         \\
\Ion{H}{2}  &   1  & $-$  & $-$ & \Ion{Ne}{8}  &   8  &  18  &  15 &         \\
\Ion{He}{1} &   1  &  25  &   0 & \Ion{Ne}{9}  &  11  &   9  &  13 &         \\
\Ion{He}{2} &  10  &  22  &  45 & \Ion{Ne}{10} &   1  &   0  &   0 &         \\
\Ion{He}{3} &   1  & $-$  & $-$ & \Ion{Mg}{9}  &   3  &  23  &   1 &         \\
\Ion{C}{4}  &   5  &  11  &   6 & \Ion{Mg}{10} &   2  &   3  &   1 &         \\
\Ion{C}{5}  &  29  &  21  &  60 & \Ion{Mg}{11} &   5  &   6  &   3 &         \\
\Ion{C}{6}  &  15  &  21  &  26 & \Ion{Mg}{12} &   1  &   0  &   0 &         \\
\Ion{C}{7}  &   1  & $-$  & $-$ & \Ion{Si}{10} &   1  &  31  &   0 &         \\
\Ion{N}{5}  &   5  &  15  &   6 & \Ion{Si}{11} &   3  &  23  &   1 &         \\
\Ion{N}{6}  &  17  &   7  &  33 & \Ion{Si}{12} &   5  &   4  &   6 &         \\
\Ion{N}{7}  &  15  &  21  &  30 & \Ion{Si}{13} &   1  &   0  &   0 &         \\
\Ion{N}{8}  &   1  & $-$  & $-$ & \Ion{S}{13}  &   9  &  21  &  10 &         \\
\Ion{O}{6}  &   5  &  30  &   6 & \Ion{S}{14}  &   9  &   1  &  15 &         \\
\Ion{O}{7}  &  19  &   7  &  32 & \Ion{S}{15}  &   1  &   0  &   0 &         \\
\Ion{O}{8}  &  15  &  30  &  30 & \Ion{IG}{14} &   6  &   0  &  20 & (21757) \\
\Ion{O}{9}  &   1  & $-$  & $-$ & \Ion{IG}{15} &   6  &   0  &  20 &  (4794) \\
            &      &      &     & \Ion{IG}{16} &   5  &   0  &  14 &   (184) \\ 
            &      &      &     & \Ion{IG}{17} &   4  &   0  &   9 &  (3409) \\
            &      &      &     & \Ion{IG}{18} &   1  &   0  &   0 &         \\ 
\hline
\end{tabular}
\end{table}

\section{Spectral analysis}
\label{sect:analysis}

A first grid of models is composed of H+He+C+N+O with solar abundance ratios \citep{AEA09} within 
\TeffwM{0.45 - 1.05} and a fixed surface gravity of \loggw{9} \sA{fig:fluxhhecno}.
We note that all synthetic energy distributions (SEDs) in our model grids 
described here are available at in Virtual Observatory (\emph{VO}) compliant form 
from the \emph{VO} service 
\emph{TheoSSA}\footnote{http://vo.ari.uni-heidelberg.de/ssatr-0.01/TrSpectra.jsp?}
provided by the \emph{German Astrophysical Virtual Observatory}
(\emph{GAVO}\footnote{http://www.g-vo.org}) as well as
atables\footnote{http://astro.uni-tuebingen.de/$^\sim$rauch/TMAF/TMAF.html}
for the use with
\emph{XSPEC}\footnote{http://heasarc.gsfc.nasa.gov/docs/xanadu/xspec}.

\begin{figure}[ht]
\begin{center}
  \resizebox{\hsize}{!}{\includegraphics[]{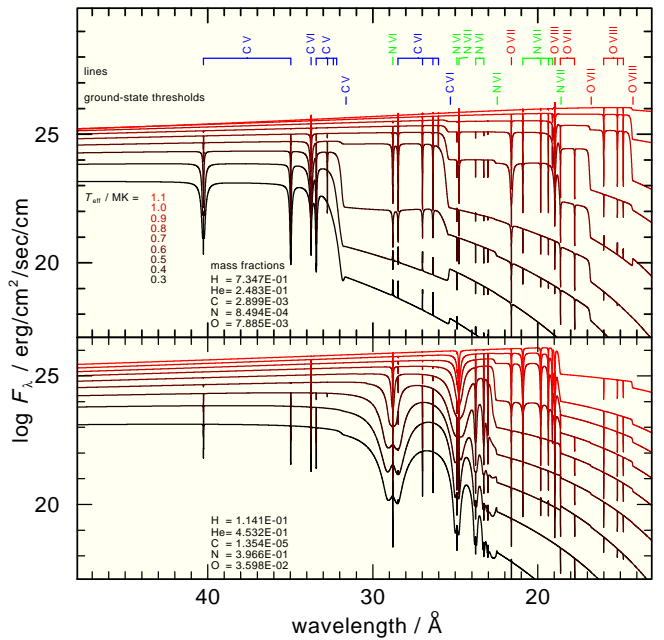}}
  \caption[]{Comparison of NLTE model fluxes (H+He+C+N+O, solar abundances, top panel) at different \Teff.
             The wavelength interval corresponds to $0.95-0.26$\,keV.
             The positions of C, N, and O lines (at top) and bound-free absorption thresholds are marked.
             In the bottom panel, the abundance ratios are taken from nova models of Prialnik (priv\@. comm.) except for 
             the carbon abundance which is reduced by a factor of 100. 
            }
  \label{fig:fluxhhecno}
\end{center}
\end{figure}

\subsection{Preliminary analysis}
\label{sect:preliminary}

 We started with the \emph{XMM-Newton} RGS spectra of 2003 April \sT{tab:obs}. The data
were reduced with the  ESA \emph{XMM-Newton} Science Analysis System\footnote{http://xmm.esa.int/sas/} (SAS)
software, version 5.3.3, using the latest calibration files available.
The RGS dispersion gratings cover the  $5 - 35$\,\AA\ wavelength range ($2.48-0.35$\,keV),
although we obtained useful signal only in the $5 - 26$\,\AA\ range ($2.48-0.48$\,keV).
The RGS data are piled-up, but in a dispersion instrument, piled-up
events at a discrete wavelength increase in pulse height amplitude by an integer
multiple of the intrinsic energy. Furthermore, since the softness of the source
precludes any intrinsic photons from higher spectral orders, we can confidently
identify events that occur within the higher order spectral masks for an on-axis
point source as piled-up first order photons. This is verified by line matching
of the piled-up events using the first order response matrix. Source events
dominate over background and scattered source light in the first three orders.
Consequently we added  events within the second and third order extraction masks
to the first order events, thus reclaiming the piled-up events and increasing
the signal-to-noise of the spectra.

\begin{table*}[ht]
\caption{X-ray observations of V4743 Sgr used in the analysis.}
\label{tab:obs}
\begin{center}
\begin{tabular}{ccllc}
\hline
\hline
\noalign{\smallskip}
                  &            &            &                     & exp\@. time \vspace{-0.7em}\\
satellite         & instrument & ObsId      & Obs date (start)    &             \vspace{-0.7em}\\
                  &            &            &                     & sec                        \\
\hline
\emph{Chandra}    & HRC+LETG   & 3775       & 2003-03-19 09:30:12 & 27058       \\
\emph{XMM-Newton} & RGS-1      & 0127720501 & 2003-04-04 22:24:12 & 35215       \\ 
\emph{XMM-Newton} & RGS-2      & 0127720501 & 2003-04-04 22:24:12 & 35214       \\ 
\emph{Chandra}    & HRC+LETG   & 3776       & 2003-07-18 21:35:29 & 13746       \\
\emph{Chandra}    & HRC+LETG   & 4435       & 2003-09-25 23:51:50 & 14378       \\
\emph{Chandra}    & HRC+LETG   & 5292       & 2004-02-28 01:07:42 & 13530       \\
\hline
\end{tabular}
\end{center}
\end{table*}

In a first step, we compared our SEDs with the \emph{XMM-Newton} observation of \vsgr\ via \emph{XSPEC}.
We let the neutral hydrogen column density $N_\mathrm{H\,I}$ [cm$^{-2}$] and \Teff\
vary as free parameters in \emph{XSPEC}.
\emph{XSPEC} determines \Teffw{709} and $\log N_\mathrm{H\,I} = 20.58$.
In \ab{fig:xspechhecno} (top panel), we show this \emph{XSPEC} fit.
The reader may miss the typical \emph{XSPEC} residuals at the bottom of the plots.
Since our models do not include all the elements with all the lines that may be
exhibited in the observation, the residuals are not that helpful like in
comparisons of simple models where continuum slope and a handful of strategic lines
has to be reproduced. Thus, we decided to omit the residual plots generally. 
Instead, the reader should judge the strengths of prominent
lines and absorption edges with respect to the (estimated) local flux continuum.

\begin{figure}[ht]
\begin{center}
  \resizebox{\hsize}{!}{\includegraphics[]{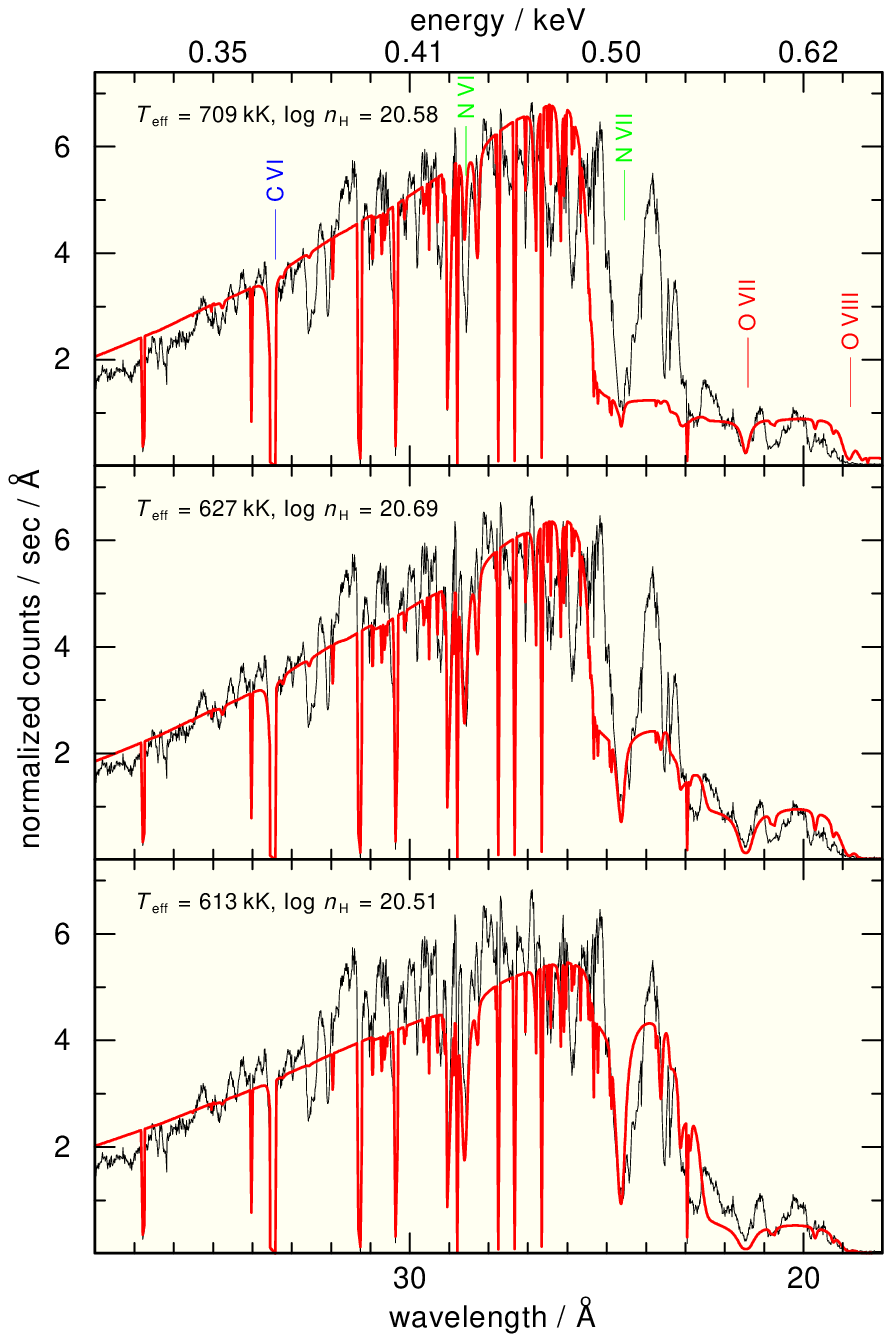}}
  \caption[]{Comparison of the \emph{XMM-Newton} RGS-1 observation of \vsgr\  \sT{tab:obs} with
             NLTE model-atmosphere fluxes (H+He+C+N+O models with solar abundances, but different carbon abundances: 
             top:    [C]\,=\,0.0, 
             middle: [C]\,=\,$-$1.0, 
             bottom: [C]\,=\,$-$2.0).
             \Teff\ and $N_\mathrm{H}$ are determined with \emph{XSPEC}. 
             The SEDs are shifted by $-1\,800\,\mathrm{km/sec}$ in order to match the observed position of the 
             \Ion{N}{7} resonance line.
             The positions of resonance lines of \Ion{C}{6}, \Ion{N}{6}, \Ion{N}{7}, \Ion{O}{7}, and \Ion{O}{8} are marked.
            }
  \label{fig:xspechhecno}
\end{center}
\end{figure}

\begin{figure}[ht]
\begin{center}
  \resizebox{\hsize}{!}{\includegraphics[]{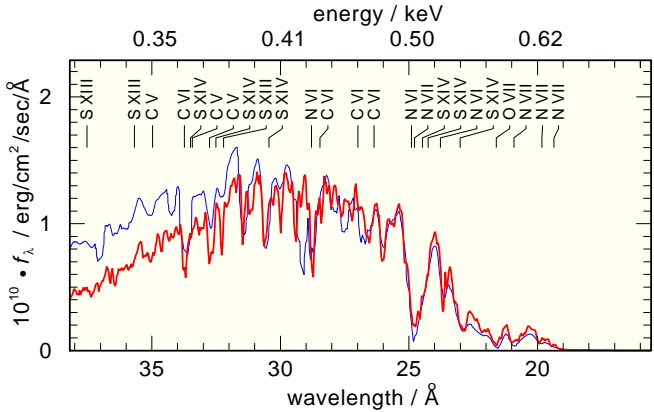}}
  \caption[]{Comparison of the RGS fluxed spectrum of \vsgr\ obtained from the \emph{XMM-Newton} pipeline (thick line) with the
             absolute flux calculated by \emph{XSPEC} (thin line, see text for details). The \emph{XSPEC} spectrum
             is convolved with a Gaussian of 0.10\,\AA\ (FWHM) in order to match the resolution of the binned
             RGS fluxed spectrum. 
             Note that both spectra agree well for $\lambda \sla 32$\,\AA.
             The marks indicate positions of lines that are included in our models.
            }
  \label{fig:abs_comp}
\end{center}
\end{figure}

As expected, a solar composition does not well represent the white dwarf spectrum in \vsgr. The flux 
at wavelengths smaller than 26\,\AA\ is far too low. In test calculations, we decreased the carbon
abundance to [C]\,=\,$-$1.0 ([X] denotes log [mass fraction / solar mass fraction] of species X) and to [C]\,=\,$-$2.0.
We achieved a much better fit at [C]\,=\,$-$2.0, \Teffw{610}, and $\log N_\mathrm{H\,I} = 20.54$
for $\lambda < 26$\,\AA\ \sA{fig:xspechhecno}. 

 We wondered whether the
small wavelength interval ($18 - 38$\,\AA) of our \emph{XMM-Newton}
RGS spectra leads in the \emph{XSPEC} fit procedure
to a smaller $N_\mathrm{H\,I}$ than found by
\citet[][$\log N_\mathrm{H\,I} = 21.60$, \Teffw{580}]{PEA05} 
obtained fitting the a \emph{Chandra} LETG spectrum ($18 - 58$\,\AA). 
In \se{sect:spevolution} we show however, that we fitted the \emph{Chandra} LETG spectrum of
March 19, 2003 and used the same wavelength range as \citet{PEA05}, deriving
\Teffw{601} and $\log N_\mathrm{H\,I} = 20.95$ for the [C]\,=\,$-$2.0 model.
 A decreasing N$_\mathrm{H\,I}$ from March to April is consistent with intrinsic absorption of
 the ejecta clearing up.

\subsection{Spectral analysis of the flux-calibrated spectrum}
\label{sect:fine}

Before we start now with a detailed analysis, 
we mention that the calibration of the instruments that
were used to obtain our data is still an issue.
Different calibration products (response matrices and effective areas)
may play an important role on the results obtained from
fits to X-ray spectra. Cross calibration between instruments
has improved over the years \citep[see, e.g\@.][]{BBR06,BBR08}
but there is yet room for further improvement.

While the wavelength scale of X-ray gratings instruments is generally
very well known, the effective areas of these instruments are less well
calibrated. For example, a comparison between the continuum calibrations
of the gratings and CCD instruments on board \emph{XMM-Newton} by 
\citet{SEA08}
shows deviations between the RGS and the EPIC-pn of 3\,\% above
0.5\,keV and about 10\,\% below that energy. While not negligible, these
deviations are still smaller than the deviations between data and model
caused by simplifications in the atmosphere modeling, and can therefore
be ignored.

The preliminary analysis presented in the last section shows clearly that the carbon abundance in \vsgr\ is strongly subsolar. 
The determination of \Teff\ and a more precise abundance determination is
hampered by the complex \emph{XSPEC} fitting procedure and the necessity to calculate extended grids of models 
with different \Teff\ if the abundances are changed. The latter results in unreasonable computational times.

  A more straightforward approach is to use the flux-calibrated spectra
 instead of count-rate spectra. However, these spectra  
contain uncertainties in the flux. We tried to calibrate
 the RGS spectra of \vsgr\ in two different ways.
 One way is by making
 use of the provided \emph{XMM-SAS} pipeline products, the so called 
``fluxed'' spectra. Such spectra are e.g\@. shown by \emph{BiRD}\footnote{http://xmm.esac.esa.int/BiRD/}, 
the Browsing Interface for RGS Data, with
 the purpose of visualizing the data free from the 
peculiarities of the instrument. However,
 RGS fluxed spectra are computed in the pipeline by dividing the extracted
 spectrum by the effective 
area calculated from its corresponding response matrix. This procedure neglects
 the redistribution of monochromatic 
response into the dispersion channels, and therefore these spectra
 should not be used for very detailed analysis.
The second version of the
flux-calibrated spectrum is calculated by \emph{XSPEC} and is not
independent from the models which are used within the \emph{XSPEC} fit procedure \sA{fig:abs_comp}.

We are aware of these uncertainties, but with the aim to determine \Teff\ 
within an error range of
about 10\,\% and element abundances within 0.5\,dex, we used the flux-calibrated
 spectra. 
We calculated synthetic profiles of the resonance lines of the two highest ionization stages of C, N, 
and O \sA{fig:CNOresonance}. These appear to be strongly dependent of \Teff\ within the parameter range of our grids.

\begin{figure}[ht]
\begin{center}
  \resizebox{\hsize}{!}{\includegraphics[]{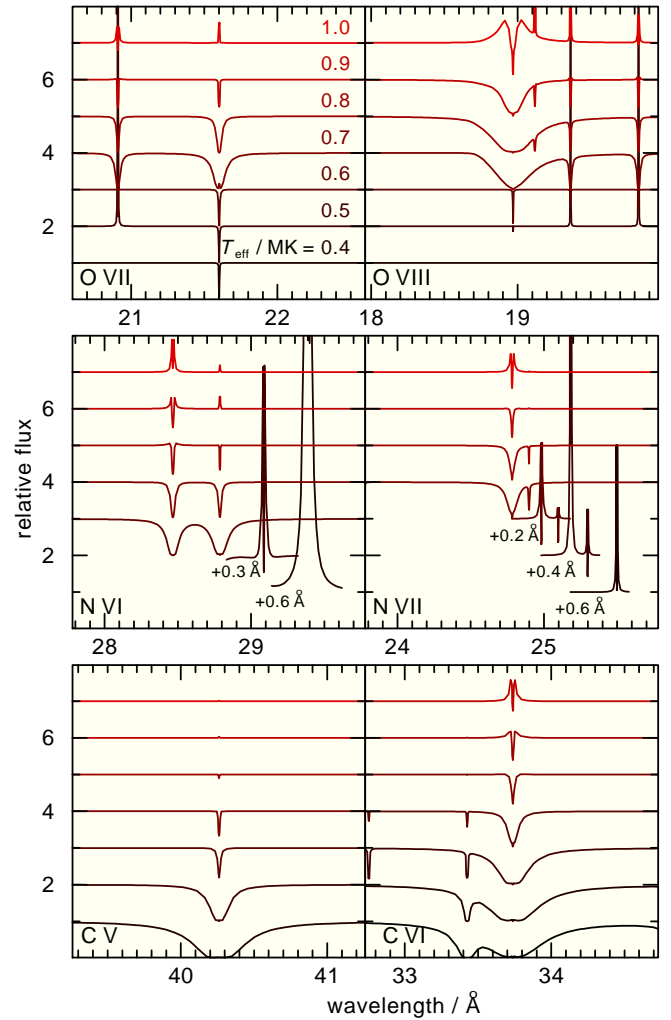}}
  \caption[]{Synthetic profiles of the \Ion{O}{7}, \Ion{O}{8}, \Ion{N}{6}, \Ion{N}{7}, \Ion{C}{5}, and \Ion{C}{6}
             resonance lines, calculated from H+He+C+N+O model atmospheres with \Teff\ = 0.4 -- 1.0\,MK, \loggw{9} and
             solar abundances. The nitrogen lines of the cooler models that appear in emission are shifted for clarity.
            }
  \label{fig:CNOresonance}
\end{center}
\end{figure}

The best \emph{XSPEC} fit of the [C]\,=\,$-$2.0 model yields \Teffw{610}. Since the maximum
flux level is not well reproduced by this fit \sA{fig:xspechhecno}, we estimate that \Teff\
of \vsgr\ may be slightly higher. Thus, we selected models with \Teffw{600 - 800} and model-A
abundances \sT{tab:abundances} for a comparison with the RGS fluxed spectrum \sA{fig:teff001}. 
The flux level below $\sla$\,30\,\AA\ is strongly dependent on \Teff\ and 
at first glance, we can achieve good agreement with
the RGS flux level at \Teffw{700} and $\log N_\mathrm{H\,I}$\,=\,20.90. Moreover, the prominent absorption lines
\Ionww{N}{6}{28.79, 24.90, 23.77}, 
\Ionww{N}{7}{24.78, 20.91, 19.82, 19.32}, and 
\Ionw{O}{7}{21.60} are well reproduced by our model.
A detailed inspection of the \Ionw{O}{8}{18.96} resonance line shows that the observed continuum flux on its
high-energy wing is much lower than in the synthetic spectrum.
A stronger \Ion{N}{7} bound-free ground-state absorption edge (18.59\,\AA)
could probably decrease this flux. Consequently, we increased the nitrogen abundance in the following models.

\begin{figure}[ht]
\begin{center}
  \resizebox{\hsize}{!}{\includegraphics[]{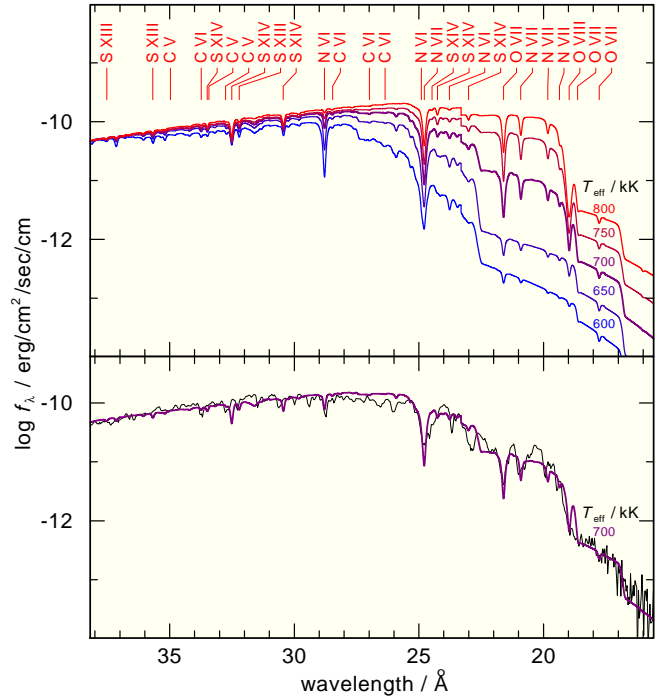}}
  \caption[]{   Top: SEDs of our model with [C]\,=\,$-$2.0 and \Teffw{600 - 800}.
                     They are normalized to the \Teffw{700} flux level at 38\,\AA\ for this comparison.
                     The positions of lines are marked.
             Bottom: Comparison of the synthetic spectra of our [C]\,=\,$-$2.0, \Teffw{700} model
                     with the RGS fluxed spectrum of \vsgr.
            
            }
  \label{fig:teff001}
\end{center}
\end{figure}

In \ab{fig:teff001} the \Teff -dependency of the strengths of prominent bound-free absorption edges (for an identification,
see \ab{fig:N_thresholds}) is shown. Especially, the \Ion{N}{6} and \Ion{N}{7} ground-state edges at
22.46\,\AA\ and 18.59\,\AA, respectively, appear very sensitive on \Teff. In order to improve the fit, we performed 
some fine-tuning of the abundances (cf\@. \ta{tab:abundances}, model B). E.g\@. in the case of sulfur, model A
\sT{tab:abundances} yields a much too strong \Ion{S}{14} line at 32.51\,\AA, i.e\@. the sulfur abundance is too high
in the relevant \Teff\ range. In the case of nitrogen, we find prominent absorption edges as well as absorption lines in the
observation which are suitable to adjust the abundance. In the case of other species, it is difficult to determine
their abundances precisely because they do not exhibit significant features in the \emph{XMM-Newton} observation. However, this can
be used to determine at least upper abundance limits. \aba{fig:upper_ten} demonstrates
how  the SEDs change if
the abundance of one particular element is artificially increased by a factor of ten. We note that only nitrogen,
oxygen, silicon, and the iron-group elements have an influence on the flux in the RGS wavelength range if their
abundances are increased.

\begin{figure}[ht]
\begin{center}
  \resizebox{\hsize}{!}{\includegraphics[]{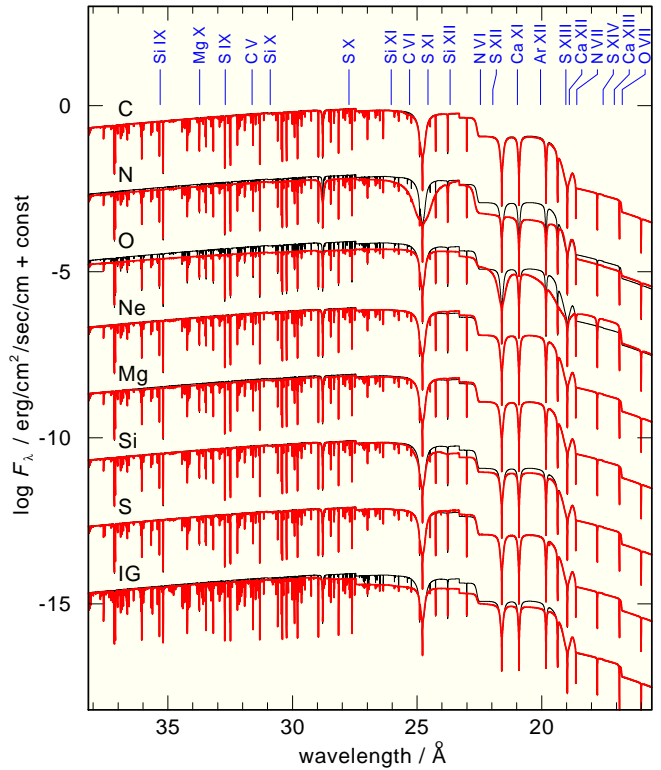}}
  \caption[]{Comparisons of the SED of our \Teffw{750} model B (thin line, \ta{tab:abundances}) to SEDs of models where
             the abundance of one element (indicated by labels) is increased by a factor of ten. The positions of ground-state
             absorption edges are marked at the top.
            }
  \label{fig:upper_ten}
\end{center}
\end{figure}

We compared the strengths of the \Ion{N}{6} and \Ion{N}{7} edges in our models with the observation \sA{fig:N_thresholds}. 
With these models (\ta{tab:abundances}, model B), we achieve the best fit at \Teffw{740}
and $\log N_\mathrm{H\,I} = 20.85$.

\begin{figure}[ht]
\begin{center}
  \resizebox{\hsize}{!}{\includegraphics[]{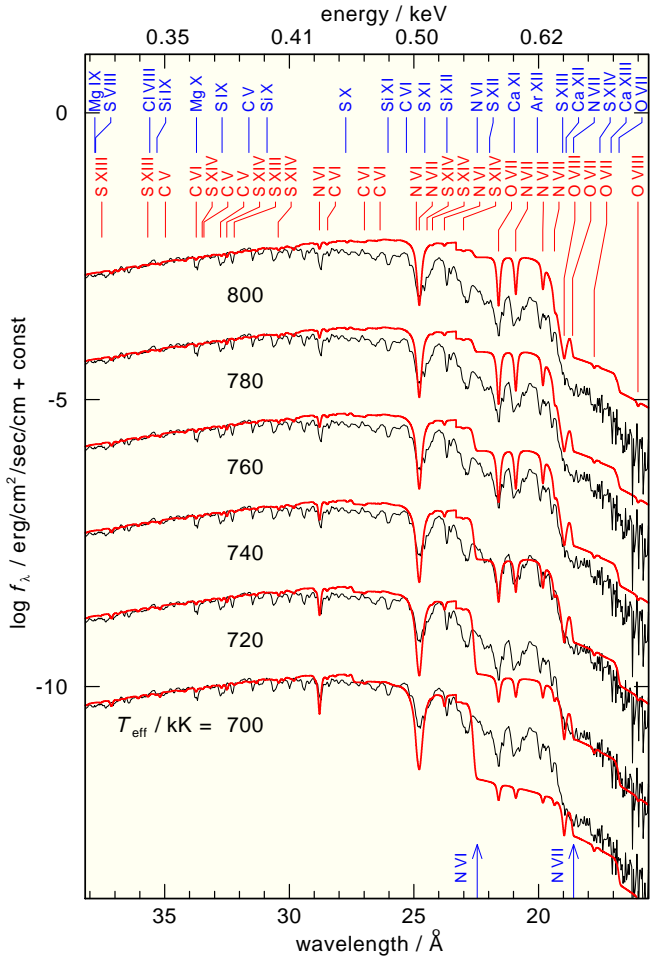}}
  \caption[]{Dependance of the strengths of the \Ion{N}{6} and \Ion{N}{7} ground-state thresholds (positions marked by
             the arrows) on \Teff.
             The SEDs are compared with the RGS fluxed spectrum of \vsgr\ and normalized to match the 
             flux at 38\,\AA. The SEDs were convolved with a Gaussian of 0.10\,\AA\ (FWHM).
             Positions of ground-state thresholds are marked at top, of those line transitions
             just below that marks.
            }
  \label{fig:N_thresholds}
\end{center}
\end{figure}

\begin{table}
\caption{Element abundances in our models given in [X] = log [mass fraction / solar mass fraction] of species X.
         Solar abundances are taken from \citet{AEA09}.
         We note that in model B, the abundances for Ne, Mg, S, and IG (Ca -- Ni) are determined upper limits.
         An error range of $\pm 0.3\,\mathrm{dex}$ is estimated from detailed comparison with the
         observations.}
\label{tab:abundances}
\begin{tabular}{lr@{.}lr@{.}l}
\hline
\hline
\noalign{\smallskip}
element & \multicolumn{2}{c}{model A} & \multicolumn{2}{c}{model B} \\
\hline
\noalign{\smallskip}
H       &                     -0&573  &                     -0&688  \\
He      &                      0&497  &                      0&382  \\
C       &                     -2&614  &                     -1&513  \\
N       &                      0&386  &                      1&803  \\
O       &                      0&425  &                      1&528  \\
Ne      &                     -0&333  &                     -0&474  \\
Mg      &                      0&000  &                     -0&454  \\
Si      &                      0&000  &                      0&167  \\
S       &                      0&000  &                     -1&583  \\
IG      &                      0&000  &                      0&828  \\
\hline
\hline
\end{tabular}
\end{table}

In \ab{fig:chandraabs}, we show a comparison of \emph{TMAP} SEDs with the flux-calibrated
\emph{Chandra} spectra from 2003. In 2004, the measured flux level and thus, \Teff\ are 
much lower and we disregard that observation here. Furthermore, for reasons of simplicity, 
we used a \Teffw{720} model with model-B abundances \sT{tab:abundances} and varied only 
the C and N abundances in order to improve the fit. The observed strengths of the 
N\,{\sc vi} and N\,{\sc vii} resonance lines as well as their ground-state edges 
are well reproduced for all observations. 
Since the N\,{\sc vi} / N\,{\sc vii} ionization balance
is very sensitive to \Teff\ \sA{fig:CNOresonance}, \Teffw{720} appears a good estimate.
Our model fits show that the N abundance is decreasing from March to September by a
factor of about ten. The C lines, e.g\@. at about 34\,\AA\ are well matched in September while
they appear too weak in March and July. A higher C abundance would result in the appearance
of a strong C\,{\sc vi} ground-state absorption edge at 25.3\,\AA\ that is not visible
in the observation.

\begin{figure}[ht]
\begin{center}
  \resizebox{\hsize}{!}{\includegraphics[]{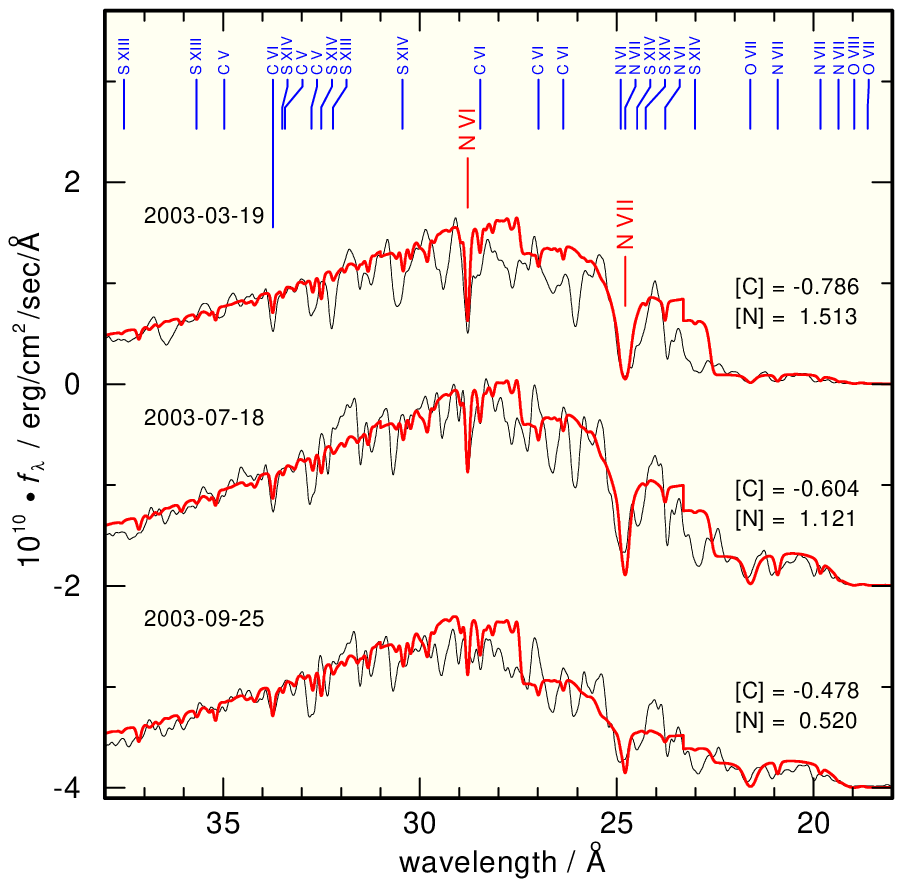}}
  \caption[]{Comparison of flux-calibrated \emph{Chandra} observations of \vsgr\  \sT{tab:obs} with
             our SEDs (\Teffw{720}, \loggw{9.0}). All abundances, but for C and N, are 
             from model-grid B \sT{tab:abundances}. For clarity, the July and September 
             observations are shifted in $f_\lambda$ by
             $-2\cdot10^{10}$ and $-4\cdot10^{10}\,\mathrm{erg/cm^2/sec/\AA}$, respectively .
             }
  \label{fig:chandraabs}
\end{center}
\end{figure}

In a last step, we used the SEDs of our model-A and model-B grids \sT{tab:abundances} for a comparison
with the \emph{XMM-Newton} observation with \emph{XSPEC} \sA{fig:xspecfinal}. 
The differences in the SEDs of the best fitting models A and B are small. If we assume the mean
\Teff, we are able to constrain the \Teff\ range of \vsgr\ to \Teffw{740\pm 70}.
The deviations of the abundance of both models \sT{tab:abundances} show the difficulty of
a precise abundance determination if only a few spectral lines are available for the analysis.

\begin{figure}[ht]
\begin{center}
  \resizebox{\hsize}{!}{\includegraphics[]{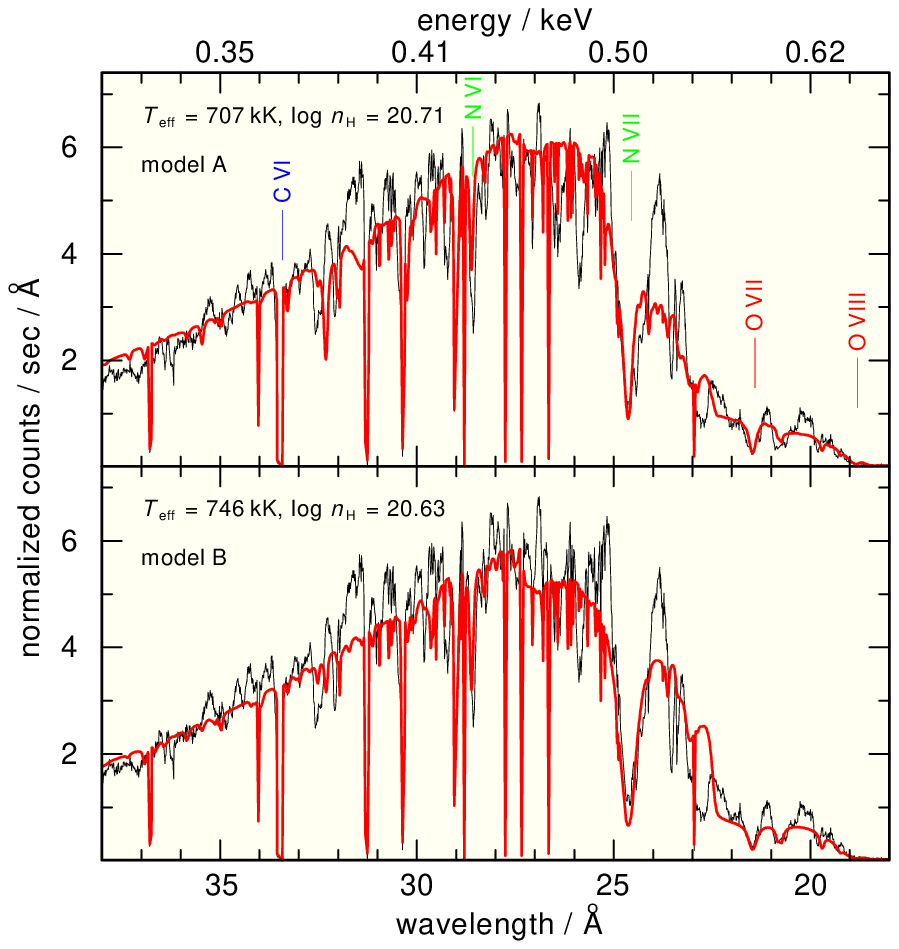}}
  \caption[]{Comparison of the \emph{XMM-Newton} RGS-1 observation of \vsgr\  \sT{tab:obs} with
             our best fitting SEDs from model grids A and B \sT{tab:abundances}.
             \Teff\ and $\log N_\mathrm{H}$ were determined with \emph{XSPEC}. 
            }
  \label{fig:xspecfinal}
\end{center}
\end{figure}

\section{The spectral evolution of V4743 Sgr}
\label{sect:spevolution}

\setcounter{footnote}{1}

We then proceeded to fit the \emph{Chandra} LETG spectra at three
 different epochs using
 the same models. The LETG spectra were
 reduced using the CIAO software\footnote{http://cxc.harvard.edu/ciao/}, version 4.1.1.

In the first 18 months after the outburst of \vsgr, \emph{Chandra} LETG spectra
were obtained roughly every three months \sT{tab:obs}. These observations allow an investigation on the 
spectral evolution, e.g\@. on the change of its \Teff. In \ab{fig:timeseries}, 
we compare the observations with SEDs of our grid (models B). No individual fine-tuning of the model abundances
for the different spectra was performed. Although this might improve the fit, we estimate that the impact
on the determined \Teff\ is small.

The \emph{Chandra} observations (energy range $0.22 - 0.69$\,keV) show that \vsgr\ was at a maximum of 
\Teff\ ($\approx 700$\,kK) for about one year, then \Teff\ appears to decrease by about 10\,\% in the 
following six months (\ab{fig:timeseries}, left panel). 

  In order to judge the quality of our spectral analysis of the Apr 2004 
\emph{XMM-Newton} observation that
covers a smaller energy range ($0.32 - 0.69$\,keV), we restricted the fit range of \emph{XSPEC}
(\ab{fig:timeseries}, right panel) for the \emph{Chandra} observations. With this restriction, the determined 
\Teff\ is in general higher and the interstellar neutral hydrogen density is smaller,
 but the deviations are 
within our expected error range of about 10\,\%.

\begin{figure*}[ht]
\begin{center}
  \resizebox{0.99\hsize}{!}{\includegraphics[]{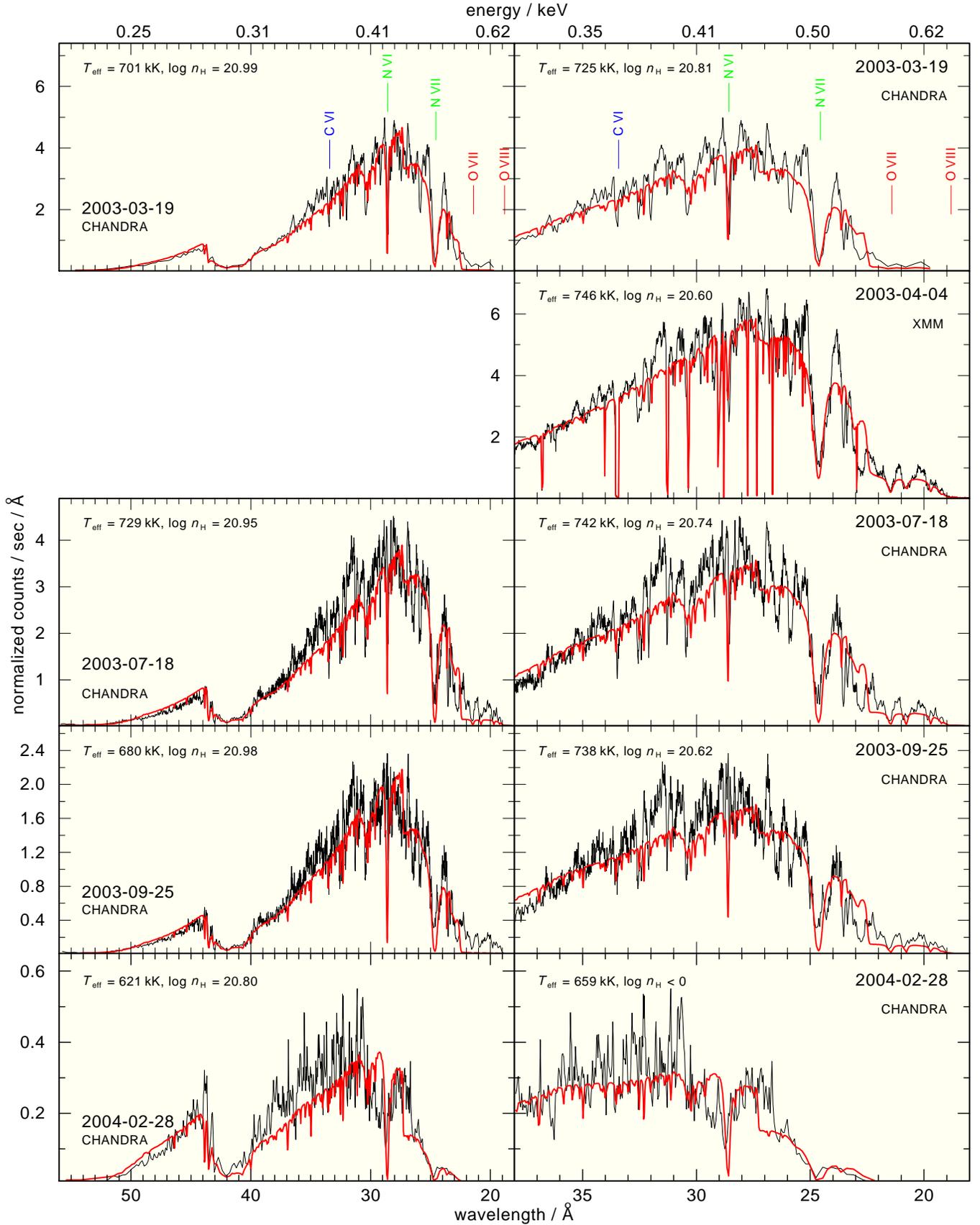}}
  \caption[]{Comparison of \emph{Chandra} and  \emph{XMM-Newton} observations of \vsgr\ \sT{tab:obs} with
             our SEDs from model grid B \sT{tab:abundances}.
             $\log N_\mathrm{H}$ and \Teff\ were determined with \emph{XSPEC}. The fit range
             was restricted to $18-56$\,\AA\ ($0.22 - 0.69$\,keV, left panel) and $18-38$\,\AA\ ($0.32 - 0.69$\,keV,
             right panel) to represent the \emph{Chandra} and  \emph{XMM-Newton} wavelength ranges, respectively.
            }
  \label{fig:timeseries}
\end{center}
\end{figure*}

\section{Results and conclusions}
\label{sect:result}

We have calculated NLTE model atmospheres including opacities
of the elements H, He, C, N, O, Ne, Mg, Si, S, and Ca -- Ni and fitted
\emph{XMM-Newton} RGS  spectra and \emph{Chandra} LETG spectra of \vsgr.

The fit to the RGS spectra \sT{tab:obs} based on these models \sT{tab:abundances}
is best at a \Teff\ of about $740\pm 70$\,kK, \loggw{9}, and $\log N_\mathrm{H} = 20.7\pm 0.3$.
Although this fit is not perfect and there are uncertainties in the so-called
RGS fluxed spectrum used for our spectral analysis,
the overall flux distribution as well as prominent line features are well in
agreement with the observation \sA{fig:xspecfinal}. The photospheric abundances have been adjusted and
we can give at least upper limits for N, O, Si, and Ca -- Ni \sA{fig:upper_ten}.
The determined C/N ratio is very low \sT{tab:abundances}, indicating that 
the material on the white dwarf surface has been processed through
 thermonuclear burning. We expect that freshly accreted material after the outburst
 or dredged-up material from the white dwarf interior, would have
 a very different range of abundances and suggest this is a proof that
 the white dwarf is retaining some accreted material after each
 outburst. This may imply that \vsgr\ is on a track towards supernova Ia
 explosion, although it retains after each outburst only just enough material
 to burn in less than 2.5 years.

 The fit to the \emph{Chandra} LETG grating spectra shows that
\vsgr\ reached its highest temperature around April 2003 and remained
at that temperature at least
until September 2003. The duration of the constant bolometric luminosity
 phase at constant \Teff\ was between 2 and 2.5 years, 
probably the average for a classical nova \citep[e.g.,][]{OEA01}.
In \citet{PEA79}, \Teff\ of
 the constant bolometric luminosity phase is 780\,000\,K for a $1.25\,\mathrm{M_\odot}$
  WD. \Teff\ of the calculations is not always published
 in papers on nova-outburst models \citep[cf\@.][]{KP94,PK95}, but published values range from
 460\,000\,K for a $0.8\,\mathrm{M_\odot}$ \citep{PEA79} to an extreme
 2\,500\,000\,K for a $1.4\,\mathrm{M_\odot}$ WD accreting at very high rate. The
 value of \Teff\ is dependent on mass-accretion rate, on the chemical
 composition and on the WD temperature at the onset of hydrogen burning,
 but generally a value of about 700\,000\,K is only reached for
 $M > 1.1\,\mathrm{M_\odot}$ (Prialnik 2009, priv\@. comm.).
To summarize, the peak \Teff\ of about 740\,kK indicates that the WD is very massive,
with $\approx 1.1 - 1.2\,\mathrm{M_\odot}$.

 To explain the apparently decreased nitrogen abundance 
 in the months after the outburst \sA{fig:chandraabs}, we conjecture
 that probably new material is already been
 accreted at this stage, while CNO burning
 at the bottom of the envelope is already proceeding at a lower
 rate before turn-off.

A crucial point for the understanding of processes during the nova outburst
may be the time-dependent prediction of surface abundances as well as the
spectral analysis of high-resolution X-ray spectra taken from outburst to 
the end of surface H burning. The inspection of available
 \emph{Chandra} and \emph{XMM-Newton} observations
of \vsgr, taken in the 18 months after the outburst has shown that 
obtaining new X-ray grating spectra of future novae every week is highly 
desirable.

In applications to novae, \emph{TMAP} still lacks some physics, 
most notably the velocity fields. 
Although \emph{TMAP} is not especially designed for the calculation 
of SEDs at extreme photospheric parameters, it is a flexible and
 robust tool for 
the determination of basic parameters like \Teff\ 
 for line identifications \citep[cf\@.][]{REA08}, 
 and to derive a reliable range of abundances in
 the white-dwarf atmosphere.

\paragraph{Acknowledgements.}
TR is supported by the German Aerospace Center (DLR) under grant 05\,OR\,0806.
This research has made use of the SIMBAD database, operated at CDS, Strasbourg, France.
MO is supported by Chandra-Smithsonian and XMM-Newton NASA grants for the data analysis.
This research has made use of software provided by the Chandra X-ray Center (CXC) in the 
application packages CIAO, ChIPS, and Sherpa.

\bibliographystyle{jphysicsB}
\bibliography{rauch.bbl}

\end{document}